\documentstyle{article}
\def\I0{\mathop{\rm I_0}}
\title{Comment on ``Voltage-Current Characteristics of the Two-Dimensional
Gauge Glass Model''}
\date{ }
\begin{document}
\maketitle

Li \cite{Li} has computed the current-voltage characteristics (CVC) 
of a disordered 
two-dimensional (2D) Josephson-junction array and claimed that his
data, presented in Fig.1 of Ref.[1],  gives evidence for a
finite-temperature phase transition in the 2D gauge-glass
model in contradiction to the result of equilibrium simulations \cite{Young}.
 His conclusion is based on the observation that the CVC is 
nonlinear for temperatures $T<0.2$. 
I have computed  a CVC of a resistively shunted 
{\bf single} 
Josephson junction in the same current and temperature range as 
Li \cite{Li}
did for the array, using the equation
\begin{equation}
V/RI_c=T/\pi[2/\pi \int _0 ^{\pi/2} d\phi \cosh(2\phi I/I_cT)\I0(2cos\phi/T)]
^{-1}\sinh(\pi I/I_cT),
\end{equation}
which is an analitical solution for the Langevin
equation for resistively shunted Josephson junction \cite{com}. Here $V$ is
the voltage $I$ is the current, $R$ is the shunt resistance,  $I_c$
is the critical current of the junction, and $\I0$ is zero order modified
Bessel function. 
Results are presented in 
Fig.1. In Fig.1(a) part of the data of Fig.1(b) is plotted, corresponding to 
the region of 
resistivities and currents studied by Li \cite{Li}. One can easily notice
a remarkable resemblance between Fig.1(a) and Fig.1 of Ref.[1]: the
CVC for $T<0.2$ is also nonlinear for a single junction. One can even collapse
 the data of Fig.1(a) on a
``scaling'' curve (Fig.2), which is just as well  as Li's
(see Fig.2 of Ref.\cite{Li}).
Is the single junction in a ``gauge-glass'' state? 
However, it is sufficient to look at Fig.1(b) to understand that 
the nonlinear CVC is just a finite-current effect.
I expect the same thing to happen for the 2D gauge glass, but 
to check this numerically would require 
enormous computation time, as there is no an analitical solution 
for Langevin equation for system of many junctions.
Fig.1 of Ref.\cite{Li} is no more  evidence of a 
finite-temperature 2D gauge-glass transition of a disordered 
Josephson-junction
array than Fig.1(a) is evidence of this ``transition'' for 
a single junction.

I am grateful to N.Akino, J.M. Kosterlitz, and J.M. Valles for useful 
conversations. This work was supported by National Science Foundation
Grant No. DMR-9222812.

\vspace{1cm}

M.V. Simkin\\
Department of Physics\\
Brown University\\
Providence, RI 02912-1843

\newpage


\begin {figure}
\caption{Resistivity $V/RI$ vs current $I/Ic$ of a single 
Josephson junction at T=0.05,
0.1,0.2,0.3,0.4(from bottom). (a) contains a part of the data of (b), 
corresponding to the region of currents and resistivities studied by Li[1]
and is remarkably similar to Fig.1 of Ref[1].}
\end {figure}
\begin {figure}
\caption{$V/RI|T-T_g|^{-z\mu}$ vs $I/I_c(|T-T_g|^{-\mu}/T$ using the data
as in Fig.1a. I used the same values of critical exponents ($\mu=2.2, z=2.2$)
 and transition temperasture ($T_g=0.15$) as Li (see Fig.2 of Ref.[1]). } 
\end {figure}
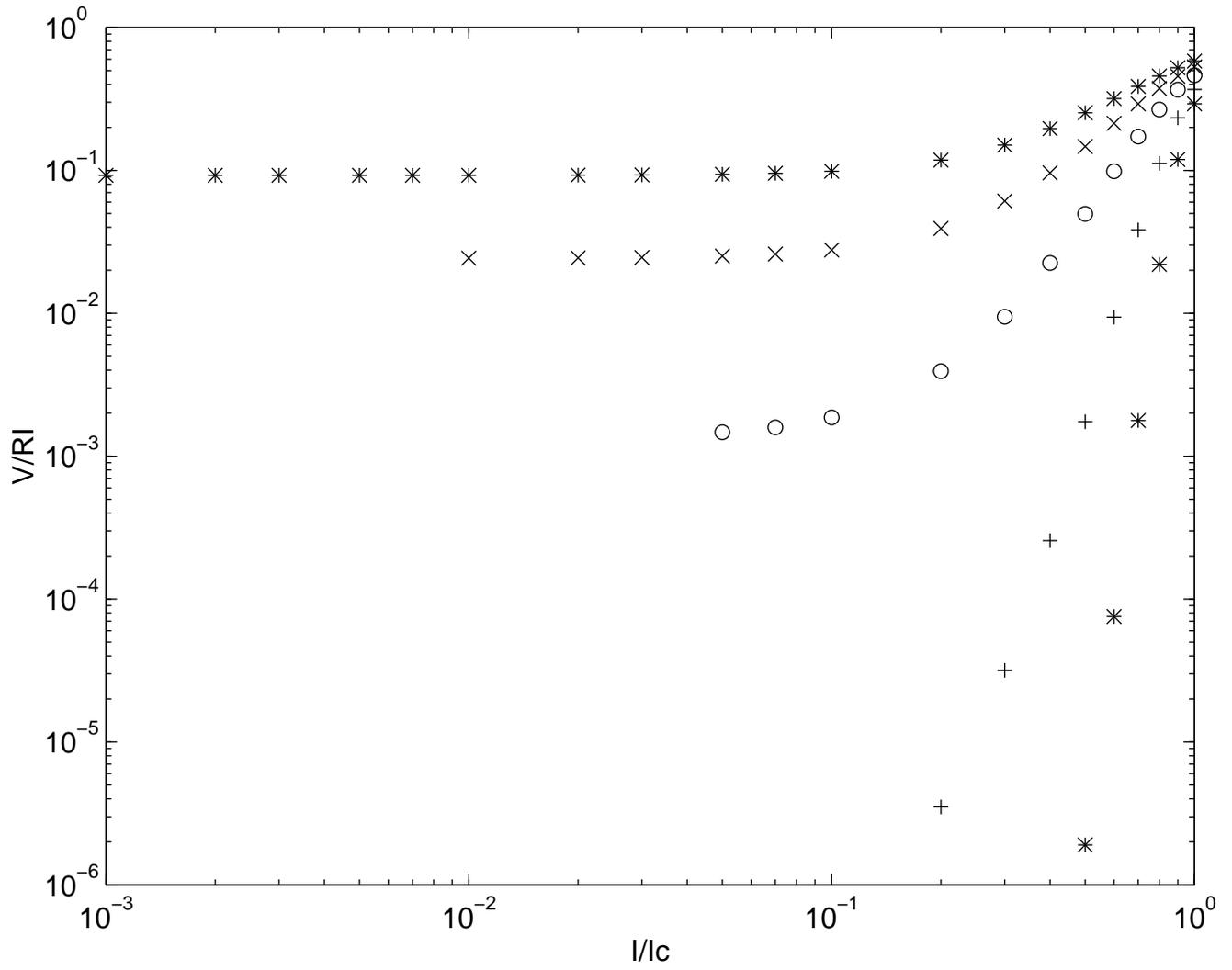
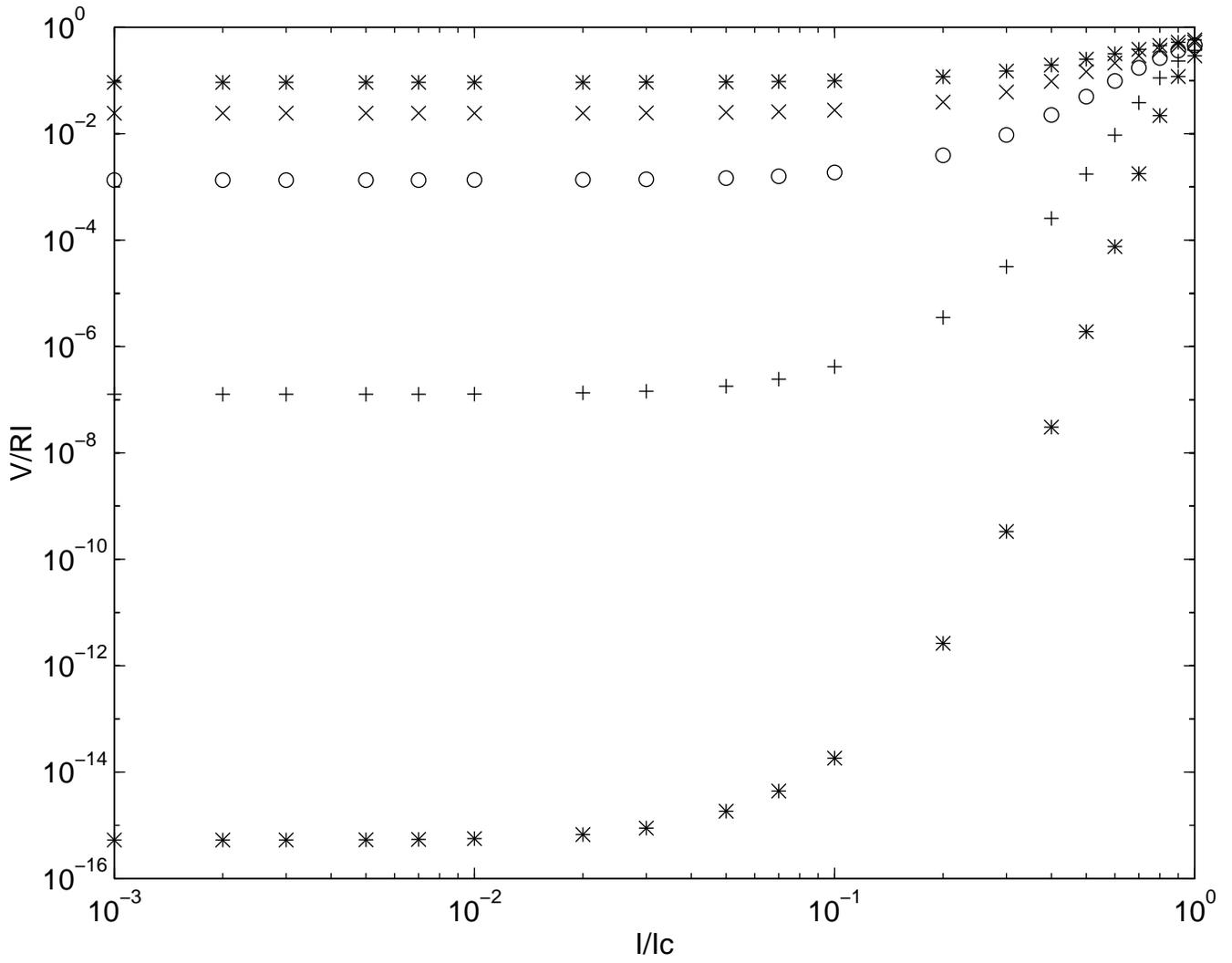
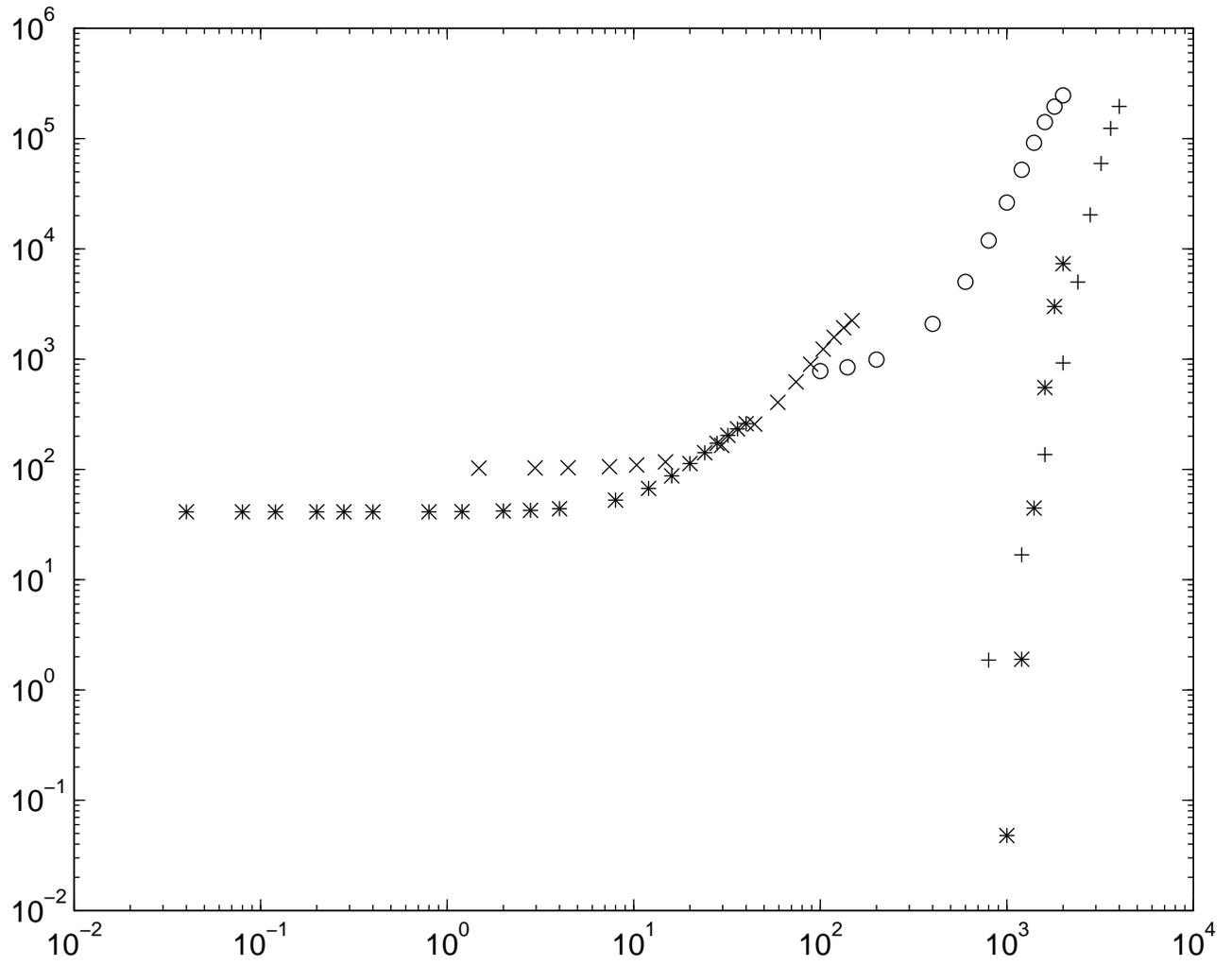
\end{document}